\documentclass[twocolumn,showpacs,aps,prl,superscriptaddress]{revtex4}

\usepackage{graphicx}
\usepackage{dcolumn}
\usepackage{amsmath}
\usepackage{epsfig}
\RequirePackage{xspace}

\newcommand{\bit}{\begin{itemize}}
\newcommand{\eit}{\end{itemize}}
\newcommand{\beq}{\begin{equation}}
\newcommand{\eeq}{\end{equation}}
\newcommand{\beqn}{\begin{eqnarray}}
\newcommand{\eeqn}{\end{eqnarray}}
\newcommand{\beqns}{\begin{eqnarray*}}
\newcommand{\eeqns}{\end{eqnarray*}}

\newcommand{\mc}{\multicolumn}

\def\rar{\rightarrow}

\def\B{\ensuremath{B}\xspace}
\def\Bz      {\ensuremath{B^0}\xspace}
\def\Bbar{\kern 0.18em\overline{\kern -0.18em B}{}\xspace}
\def\babar{\mbox{\slshape B\kern-0.1em{\smaller A}\kern-0.1em
    B\kern-0.1em{\smaller A\kern-0.2em R}}}
\def\pep2{PEP-II}
\def\invfb   {\ensuremath{\mbox{\,fb}^{-1}}\xspace}
\def\CP{\ensuremath{C\!P}\xspace}
\mathchardef\Upsilon="7107
\def\Y#1S{\ensuremath{\Upsilon{(#1S)}}\xspace}

\def\FourS {\Y4S}

\def\rpn{\rho^\pm\pi^0}
\def\rpc{\rho^0 \pi^\pm}
\def\rzp{\rho^0 \pi^0}

\def\btorp{\Bz \rar \rho\pi}

\def\bchtorpch{B^{+} \rar \rho^0 \pi^{+}}
\def\bchtorchp{B^{+} \rar \rho^{+} \pi^0}
\def\btorp{B^{0} \rar \rho^0 \pi^0}
\def\KS    {\ensuremath{K^0_{\scriptscriptstyle S}}\xspace}
\def\piz   {\ensuremath{\pi^0}\xspace}
\def\Dzb     {\ensuremath{\Dbar^0}\xspace}
\def\Dbar    {\kern 0.2em\overline{\kern -0.2em D}{}\xspace}

\def\dE{{\Delta E}}

\def\mes{\ensuremath{m_{\rm ES}}}

\def\cat{k}
\def\cont{{q\bar q}}
\def\MeVc2{${\rm MeV}/c^2$}
\def\GeVc2{${\rm GeV}/c^2$}
\newcommand{\mev}{\ensuremath{\mathrm{\,Me\kern -0.1em V}}\xspace}
\newcommand{\gevcc}{\ensuremath{{\mathrm{\,Ge\kern -0.1em V\!/}c^2}}\xspace}
\newcommand{\gev}{\ensuremath{\mathrm{\,Ge\kern -0.1em V}}\xspace}

\def\babar{$\mbox{\sl B\hspace{-0.4em} {\small\sl A}\hspace{-0.37em} 
\sl B\hspace{-0.4em} {\small\sl A\hspace{-0.02em}R}}$}
\def\rs{\raisebox{1.5ex}[-1.5ex]}

\def\deltat{\ensuremath{{\rm \Delta}t}\xspace}
\def\ee{$e^+e^-$}

\def\NN{{\rm NN}}

\def\Ampbar{\kern 0.18em\overline{\kern -0.18em A}{}_{\rho\pi}}
\def\a{\rm a}

\def\ee{\ensuremath{e^+e^-}}

\newcommand{\jprlBase}       {Phys.\ Rev.\ Lett.\xspace}
\newcommand{\jprl}      [1]  {\jprlBase\ {\bf #1}}
\newcommand{\progtp}    [1]  {{Prog.\ Th.\ Phys.\ {\bf #1}}}
\def\ea{{\em et al.}}

\def\mes {\ensuremath{m_{ES}}\xspace}

\def\dt {\ensuremath{\Delta t}}

\newcommand{\BABARPubYear}    {03}
\newcommand{\BABARPubNumber}  {037}

\newcommand{\SLACPubNumber}   {10236}

\def\figurebox#1#2#3{%
    \def\arg{#3}%
    \ifx\arg\empty
    {\hfill\vbox{\hsize#2\hrule\hbox to #2{\vrule\hfill\vbox to #1{\hsize#2\vfill}\vrule}\hrule}\hfill}%
    \else
    {\hfill\epsfbox{#3}\hfill}%
    \fi}

\begin{document}

\begin{flushleft}
\babar-PUB-\BABARPubYear/\BABARPubNumber\\
SLAC-PUB-\SLACPubNumber\\[10mm]
\end{flushleft}

\title{
{\large \bf \boldmath
Measurement of Branching Fractions and 
Charge Asymmetries \\
in \boldmath$B^\pm\to\rho^\pm \pi^0$ and
\boldmath$B^\pm\to\rho^0 \pi^\pm$ Decays, and
Search for \boldmath$B^0\to\rho^0 \pi^0$} }

%
\author{B.~Aubert}
\author{R.~Barate}
\author{D.~Boutigny}
\author{F.~Couderc}
\author{J.-M.~Gaillard}
\author{A.~Hicheur}
\author{Y.~Karyotakis}
\author{J.~P.~Lees}
\author{P.~Robbe}
\author{V.~Tisserand}
\author{A.~Zghiche}
\affiliation{Laboratoire de Physique des Particules, F-74941 Annecy-le-Vieux, France }
% ...
\collaboration{The \babar\ Collaboration}
\noaffiliation

\date{\today}

\begin{abstract}
We present measurements of branching fractions and 
charge asymmetries in \B-meson decays to $\rho^+ \pi^{0}$, 
$\rho^{0}\pi^+$ and $\rho^0\pi^0$. 
The data sample comprises
 $89 \times 10^6$ $\FourS \to B\Bbar$ decays
collected with the \babar\ detector at the \pep2 asymmetric-energy 
$B$~Factory at SLAC. We find the charge-averaged
branching fractions
${\cal B}( B^{+}\rightarrow \rho^{+}\pi^0) = (10.9 \pm 1.9{\rm (stat)}
\pm 1.9{\rm (syst)})\times
10^{-6}$ and ${\cal B}( B^{+} \rightarrow \rho^0 \pi^{+}) = (9.5 \pm
1.1 \pm 0.8) \times 10^{-6}$, and
we set a $90\%$ confidence-level upper limit
${\cal B}( \B^0 \rightarrow \rho^0\pi^0) < 2.9 \times 10^{-6}$.
We measure the charge asymmetries
$A_{CP}^{\rho^{+}\pi^0} = 0.24 \pm 0.16 \pm 0.06$ and
$A_{CP}^{\rho^0\pi^{+}} = -0.19 \pm 0.11 \pm 0.02$.
\end{abstract}

\pacs{13.25.Hw, 11.30.Er, 12.15.Hh}

\maketitle


The study of \B-meson decays into charmless hadronic final states
plays an important role in the understanding of \CP
violation in the \B system. Recently, the \babar\  experiment 
performed a search for \CP-violating asymmetries in neutral
$B$ decays to $\rho^\pm\pi^\mp$ final states~\cite{bib:350PRL}, where the
mixing-induced \CP asymmetry is related to the angle
$\alpha \equiv \arg\left[-V_{td}^{}V_{tb}^{*}/V_{ud}^{}V_{ub}^{*}\right]$
of the Unitarity Triangle~\cite{unitarity}. 
The extraction of $\alpha$ from $\rho^\pm\pi^\mp$ is complicated
by the interference of decay amplitudes with differing weak 
and strong phases.
One strategy to overcome this problem is to perform an SU(2) analysis
that uses all $\rho\pi$ final states~\cite{bib:Nir}. Assuming
isospin symmetry, the angle $\alpha$ can be determined free of 
hadronic uncertainties from a pentagon relation formed in the complex plane by the five decay amplitudes $B^0\rar\rho^+\pi^-$,
$B^0\rar\rho^-\pi^+$, $B^0\rar\rho^0\pi^0$, $B^+\rar\rho^+\pi^0$ and
$B^+\rar\rho^0\pi^+$~\cite{footn}.
These amplitudes can be determined from measurements of the 
corresponding decay rates and \CP-asymmetries. The branching fractions
have been 
measured for $B^0\rar\rho^+\pi^-$ and $B^+\rar\rho^0\pi^+$, and an upper 
limit has been set for $B^0\rar\rho^0\pi^0$ ~\cite{bib:350PRL,bib:cleo_rhopi}.

In this letter we present measurements of the 
branching fractions of the decay modes 
$B^+\rar\rho^+\pi^0$ and $B^+\rar\rho^0\pi^+$,
and a search for the decay $B^0\rar\rho^0\pi^0$.  
All three analyses follow a quasi-two-body approach~\cite{roy,bib:350PRL}.
For the charged modes we also measure the
charge asymmetry, defined as
\begin{equation}
\label{equ:Acp}
        A_{CP} \equiv
        \frac{\Gamma(B^- \rightarrow f) \,-\, \Gamma(B^+ \rightarrow \overline f)}
         {\Gamma(B^- \rightarrow f) \,+\, \Gamma(B^+ \rightarrow \overline f)}\;,
\end{equation}
where $f$ and $\overline f$ are the final state and its charge-conjugate, respectively.


The data used in this analysis were collected
with the \babar\ detector~\cite{bib:babarNim}
at the \pep2 asymmetric-energy $e^+e^-$ storage ring at SLAC.
The sample consists of $(88.9\pm1.0)\times10^{6}$ $B\Bbar$ pairs
collected at the \FourS resonance (``on-resonance''),
and an integrated luminosity of 9.6~\invfb collected about 
40~\mev below the \FourS (``off-resonance'').


Each signal \B candidate is reconstructed from three-pion 
final states that must be $\pi^{+}\pi^0\pi^0$,
$\pi^{+}\pi^-\pi^+$, 
or $\pi^{+}\pi^{-}\pi^0$. Charged tracks must have ionization-energy loss 
and Cherenkov-angle signatures 
inconsistent with those expected for electrons, kaons, protons, or
muons~\cite{bib:babarNim}. 
The $\pi^0$ candidate must have a mass that satisfies 
$0.11<m(\gamma\gamma)<0.16\gevcc$, where each photon is required to have 
an energy greater than $50\mev$ in the laboratory frame and to exhibit
a lateral profile of energy deposition in the electromagnetic calorimeter
consistent with an 
electromagnetic shower~\cite{bib:babarNim}. 
The mass of the reconstructed $\rho$ candidate must satisfy
$0.4<m(\pi^{+}\pi^0)<1.3\gevcc$ for $\rho^+$ and
$0.53<m(\pi^+\pi^-) < 0.9\gevcc$ for $\rho^0$. The tight
upper $m(\pi^+\pi^-)$ cut at 0.9$\gevcc$ is to remove contributions from
the scalar $f_0(980)$ resonance, and the tight lower cut is to 
reduce the contamination from $\KS$ decays.
To reduce contributions from $B^0\to\rho^+\pi^-$ decays, a 
$\btorp$ candidate is rejected if $0.4<m(\pi^{\pm}\pi^0)<1.3\gevcc$.
For the $\bchtorchp$ and $\btorp$ modes, the invariant mass of 
any charged track in the event and the $\piz$ must be less than 
$5.14\gevcc$ to 
reject $\B^+\rar\pi^+\pi^0$ background.
For the $\bchtorpch$ mode, we remove background from 
charmed decays $B\rightarrow \Dzb X$, $\Dzb\rightarrow K^{+}\pi^{-}$ 
or $\pi^{+}\pi^{-}$, by requiring the masses $m(\pi^+\pi^-)$ and
$m(K^+\pi^-)$ to be less than $1.844\gevcc$ or greater than $1.884\gevcc$.
We take advantage of the helicity structure of $B \rar \rho \pi$ decays
by requiring that $|\cos\theta_{\rho}|>0.25$, where $\theta_{\rho}$ is the
angle between the $\pi^0$ ($\pi^+$) momentum from the 
$\rho^{+}$ $(\rho^{0})$ decay
and the \B momentum in the $\rho$ rest frame.

Two kinematic variables, $\dE$ and $\mes$, allow the discrimination 
of signal \B decays from random combinations of tracks and
$\pi^0$ candidates. The energy difference,
$\dE$, is the difference between the $\ee$ center-of-mass (CM)
energy of the \B candidate and $\sqrt{s}/2$, where $\sqrt{s}$ is the 
total CM energy. The beam-energy-substituted mass, $\mes$, is defined by
$\sqrt{(s/2+{\mathbf {p}}_i\cdot{\mathbf{p}}_B)^2/E_i^2-{\mathbf {p}}_B^2},$
where the $B$ momentum, ${\mathbf {p}}_B$, and the four-momentum of the 
initial state 
($E_i$, ${\mathbf {p}}_i$) are measured in the laboratory frame.
For $\bchtorpch$ we require that
$-0.05< \dE < 0.05\gev$ while for both modes
containing a $\pi^0$ we relax this requirement to 
$-0.15<\dE< 0.10\gev$.
For both $\bchtorpch$ and $\btorp$ we require that $5.23<\mes<5.29\gevcc$
while for $\bchtorchp$ it is relaxed to $5.20 < \mes < 5.29\gevcc$ 

Continuum $e^+e^-\to q\bar{q}$ ($q = u,d,s,c$) events are the dominant 
background. To enhance discrimination between signal and continuum, 
we use neural networks (NN) to combine six discriminating 
variables: the reconstructed $\rho$ mass,  $|\cos\theta_{\rho}|$, 
the cosine of the angle between the \B momentum and the 
beam direction in the CM frame, 
the cosine of the angle between 
the \B thrust axis and the beam direction in the CM frame, and
the two event-shape variables that are used in the Fisher
discriminant of Ref.~\cite{bib:babarsin2b}. The event shape variables
are sums over all particles $i$ of $p_i\times|cos\theta_i|^n$, 
where $n=0$ or $2$
and $\theta_i$ is the angle between momentum $i$ and the $B$ thrust axis.
The NN for each analysis weighs the discriminating variables 
differently, according to training on  
off-resonance data and the relevant Monte Carlo (MC) simulated signal 
events. The final $\rho\pi$ candidate samples are selected with cuts
on the corresponding NN outputs.

To further discriminate further between signal and continuum  
background, for the $\btorp$ mode, we use the separation between the vertex
of the reconstructed $B$ and the vertex reconstructed for the remaining
tracks. This separation is related to $\deltat$, the difference between
the two decay times, by $\Delta z = c\beta\gamma \deltat$, where for
\pep2 the boost is $\beta\gamma=0.56$.

Approximately $33\%$, $7\%$, and $8\%$ of the events have more 
than one candidate satisfying the selection in the $\bchtorchp$, 
$\bchtorpch$, and $\btorp$ decay mode, respectively. In such cases we 
choose the candidate with the reconstructed $\rho$  mass
closest to the nominal value of $0.77\gevcc$.
Table~\ref{tab:sumtab} summarizes the numbers of events selected from
the data sample and the signal efficiencies estimated from MC simulation.
Some of the actual signal events are misreconstructed; this is primarily 
due to the presence of random combinations involving low momentum pions.
For the charged \B modes we distinguish misreconstructed
signal events with correct charge assignment from those with incorrect 
charge assignment. These numbers, estimated from MC, are also listed
in Table~\ref{tab:sumtab}.

\begin{table}[t]
\caption{Numbers of selected events from on-resonance data, 
signal efficiencies,
relative fraction of misreconstructed and wrong charge events from MC.}
{\small
\begin{center}
\setlength{\tabcolsep}{0.285pc}
\begin{tabular}{lc c c}
\hline
\hline
&&&\\[-0.3cm]
 & $\bchtorchp$ & $\bchtorpch$ & $\btorp$ \\
\hline
Selected events   & 13177 & 8551 & 7048 \\
Signal efficiency & $17.5\pm0.1\%$ & $28.3\pm0.1\%$ & $20.0\pm0.1\%$ \\
Misreconstructed  & $38.6\pm0.2\%$ & $7.1\pm0.1\%$ & $9.1\pm0.2\%$ \\
Wrong charge      & $8.1\pm0.1\%$ & $1.6\pm0.1\%$ & - \\ 
\hline \hline
\end{tabular}
\end{center}
}
\label{tab:sumtab}
\end{table}


We use  MC-simulated events to study the background
from other $B$~decays, (\B-background), which include both 
charmed ($b \to c$) and charmless decays.
In the selected $\rho^+\pi^0$ 
($\rho^0\pi^+$, $\rho^0\pi^0$)
sample we expect
$205\pm46$ ($73\pm19$, $59\pm18$) $b \to c$ and
$228\pm77$ ($92\pm11$, $74\pm22$) charmless background events.
All the three analyses share the major \B-background modes:
$B^0\to\rho^+\pi^-$, longitudinally polarized
$B^0\to\rho^+\rho^-$, and $B^+\to\rho^+\rho^0$.
Other important modes include  $\bchtorchp$ (for $\btorp$), 
$B^+\to(\a_1\pi)^+$ (for $\bchtorchp$), 
$B^+\to K^{*}(892)^{0}\pi^+$ (for $\bchtorpch$),
and background modes containing higher kaon resonances.


An unbinned maximum likelihood fit is used for each analysis
to determine event yields and charge asymmetries.
To enhance discrimination between signal and background events, we use
the $B$-flavor-tagging algorithm developed for the \babar\ measurement of
the $CP$-violating amplitude $\sin2\beta$ ~\cite{bib:babarsin2b}, 
where events are
separated into categories based on the topology of the event and the
probability of misassigning the $B$-meson flavor.
The likelihood for the $N_\cat$ candidates tagged in category 
$k$ is
\begin{equation}
\label{eq:pdfsum}
{\cal L}_k = e^{-N^{\prime}_\cat}\!\prod_{i=1}^{N_\cat}
\bigg\{ N^{\rho \pi} 
\epsilon_\cat {\cal P}_{i,\cat}^{\rho \pi}
+ N_\cat^{\cont} {\cal P}_{i,\cat}^{\cont} 
 + \sum_{j=1}^{N_B} {\cal L}^{\B}_{ij, \cat}\bigg\}\;,
\end{equation}
where $N^{\rho \pi}$ is the number of signal events 
in the entire sample, $\epsilon_\cat$ is the 
fraction of signal events tagged in category $\cat$, 
$N^{\cont}_\cat$ is the number of continuum background 
events that are tagged in 
category~$\cat$, and $N_B$ is the number of \B-background modes.
$N^{\prime}_\cat$ is the sum of the expected event yields for
signal ($\epsilon_\cat N^{\rho \pi}$), continuum 
($ N_\cat^{\cont}$) and fixed 
\B background. For the charged modes the asymmetries are introduced
by multiplying the signal yields by $\frac{1}{2}(1-Q_{i}A_{CP})$, where
$Q_i$ is the charge of $B$-candidate $i$.
The likelihood term 
${\cal L}^{\B}_{ij,\,\cat}$ corresponds to 
the $j_{th}$ \B-background contribution of the $N_B$ 
\B-background classes. The total likelihood is the product 
of likelihoods for each tagging category.

The probability density functions (PDF) for signal and continuum,
${\cal P}_{\cat}^{\rho\pi}$ and ${\cal P}_{\cat}^{q\bar q}$,
are the products of the PDFs of the discriminating variables.
The signal PDFs are given by
${\cal P}^{(\rho\pi)^+}_\cat \equiv {\cal P}^{(\rho\pi)^+}(\mes)
\cdot {\cal P}^{(\rho\pi)^+} (\dE)\cdot {\cal P}^{(\rho\pi)^+}_\cat (\NN)$ 
for the charged \B decay modes, and by
${\cal P}^{\rho^0\pi^0}_\cat \equiv {\cal P}^{\rho^0\pi^0}(\mes)
\cdot {\cal P}^{\rho^0\pi^0} (\dE)\cdot {\cal P}^{\rho^0\pi^0}_\cat (\NN)
\cdot {\cal P}^{\rho^0\pi^0}_\cat (\deltat)$ for $\Bz \to \rho^0\piz$.
Each signal PDF is decomposed into two
parts with distinct distributions: signal events that are correctly 
reconstructed and signal events that are misreconstructed.
For the charged \B modes, each PDF for the 
misreconstructed events is further divided into a right-charge and 
wrong-charge part.
The $\mes$, $\dE$, and NN PDFs for signal and for \B background 
are taken from MC simulation.
For continuum, the yields and PDF parameters are determined simultaneously
in the fit to on-resonance data.

In the $\btorp$ decay the $\dt$ distributions for 
signal and \B background are modeled from fully reconstructed $B^0$ 
decays from data control samples~\cite{bib:babarsin2b}. 
The continuum $\deltat$ parameters are free in the fit to on-resonance
data.

\begin{table}[t]
\caption{Summary of the systematic uncertainties.}
{\small
\begin{center}
\setlength{\tabcolsep}{0.305pc}
\begin{tabular}{lc c c c c }
\hline
\hline
 & $\rho^+ \pi^0$ & $\rho^0 \pi^+$ &  $\rho^0 \pi^0$
 & \mc{1}{ c }{$A_{CP}^{\rho^+\pi^0}$} & $A_{CP}^{\rho^0\pi^+}$   \\ 
\rs{Error source} & \mc{3}{ c }{(events)}
 &  \mc{2}{ c }{($10^{-2}$)} \\ \hline
Signal model  & 10.7 & 3.8 & 3.3 & 3.4 & 0.3 \\
Fit procedure bias & 14.4 & 8.2 & 2.0 & - & - \\
\B background  & 11.2 & 2.3 & 3.3 & 5.0 & 2.2 \\
Detector charge bias & - & - & - & 1.0 & 0.9 \\
\hline
Total fit error  & 21.1 & 9.3 & 5.1 & 6.1 & 2.4 \\
\hline
Relative efficiency error  &11.6$\%$  &7.2$\%$ &7.0$\%$ & - & - \\
\hline \hline
\end{tabular}
\end{center}
}
\label{tab:sys_table}
\end{table}

To validate the fit procedure,
we perform fits on large MC samples that contain the measured
number of signal and continuum events and the expected
\B-background.
Biases observed in these tests are largely due to 
correlations between the discriminating variables, which are not 
accounted for in the PDFs.
For $\rho^+ \pi^0$ and $\rho^0 \pi^+$
they are not negligible and are used to correct
the fitted signal yields. In addition, the full fit biases
are assigned as systematic uncertainties on all three signal
yields.

Contributions to the systematic errors
are summarized in Table~\ref{tab:sys_table}.
Uncertainties in the signal MC simulation
are obtained from a topologically similar control sample of fully reconstructed
$B^{0} \rightarrow D^{-} \rho^{+}$ decays. For the $\bchtorchp$ channel
we also use $B^{+} \rightarrow K^{+} \pi^{0}$ decays to estimate the 
uncertainty in the $\dE$ model. We vary the signal parameters, that are fixed 
in the fit, within their estimated errors and assign the effects on the signal 
yields and charge asymmetries as systematic errors.
The expected yields from the \B-background modes are varied according to
the uncertainties in the measured or estimated branching fractions.
Since \B-background modes may exhibit direct \CP violation, the
corresponding charge asymmetries are varied within their physical ranges.
For $B^0\to\rho^0\pi^0$, the systematic uncertainty due to interference
with $B^0\to\rho^+\pi^-$ is found to be 1.5 events. This is obtained
by repeating the fit to data, after removing the cut on $m(\pi^{\pm}\pi^0)$.
Systematic errors due to possible nonresonant 
$B^0\to\pi^+ \pi^-\pi^0$ decays are derived 
from experimental limits~\cite{bib:cleo_rhopi}.
Contributions from nonresonant $B^+\rightarrow \pi^+\pi^0\pi^0$
for the $\rho^+\pi^0$ mode and $\B^+ \rightarrow \pi^+\pi^-\pi^+$
for the $\rho^0\pi^+$ mode are estimated to be negligible. 
For the $\bchtorpch$ and $\btorp$ decay modes,
systematic uncertainties due to interference between $\rho^0$ and
$f_0(980)$ or a possible broad scalar $\sigma(400-1200)$ were also 
studied and found to be negligible. 
Repeating the selection and fit for all three	
modes, without using the $\rho$-candidate mass and helicity angle, gives 
results that are compatible with those reported here.
In the $\bchtorpch$ case, the analysis was repeated in the region
$|\cos\theta_{\rho}|<0.25$,
and the resulting signal yield was consistent with zero.

\begin{figure}[t]
  \centerline{ 	\epsfxsize4.5cm\epsffile{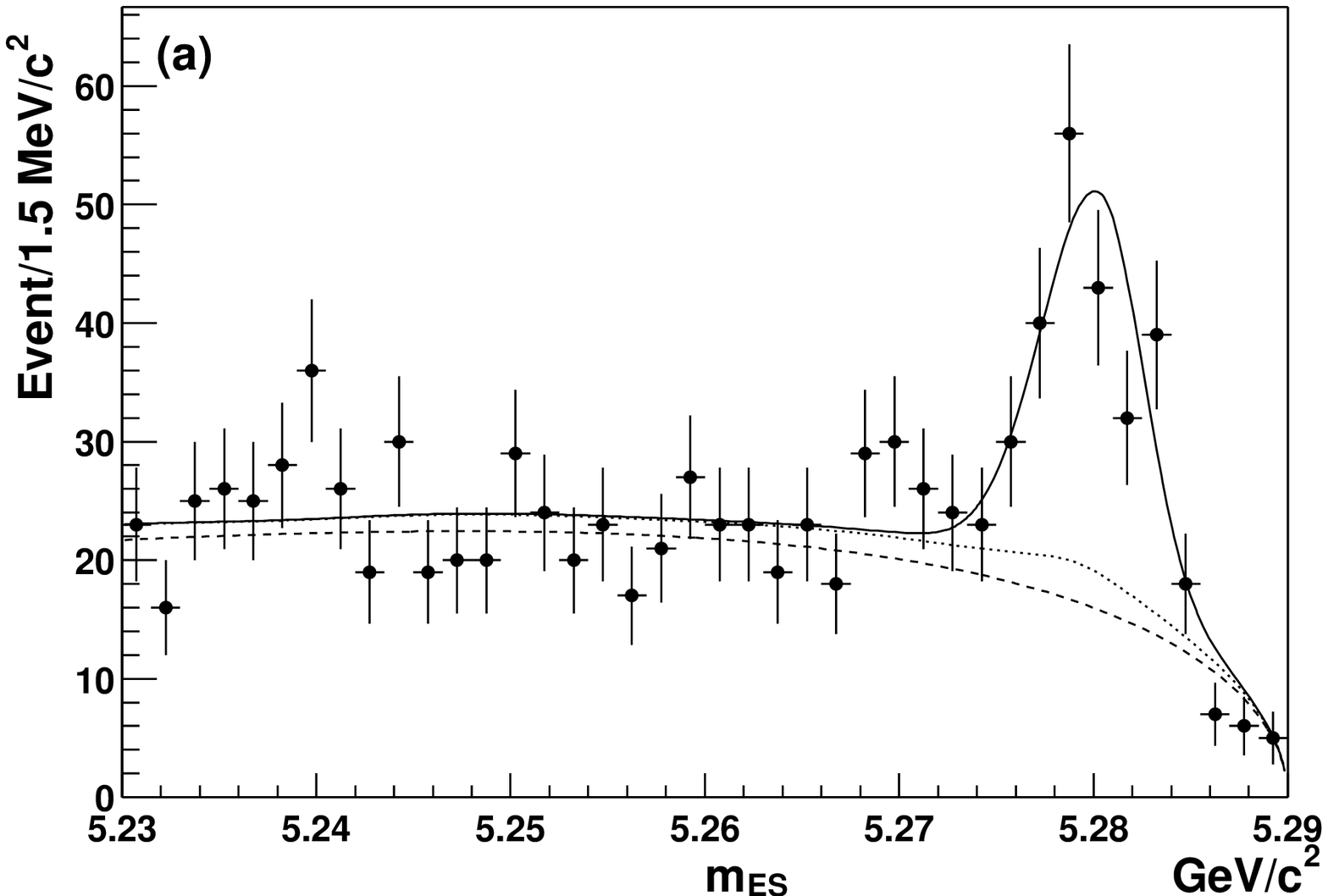}
		\epsfxsize4.5cm\epsffile{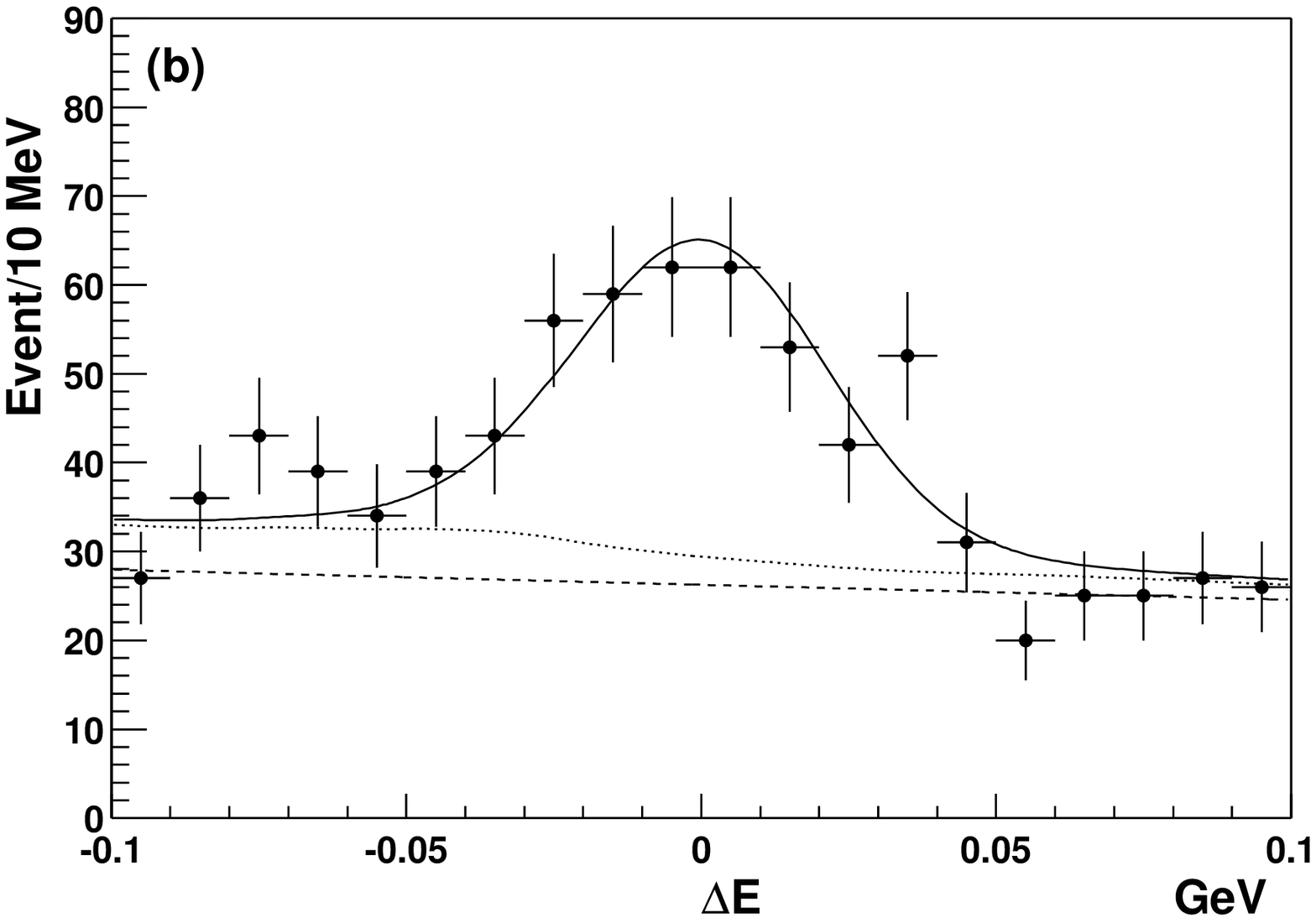}}
  \centerline{ 	\epsfxsize4.5cm\epsffile{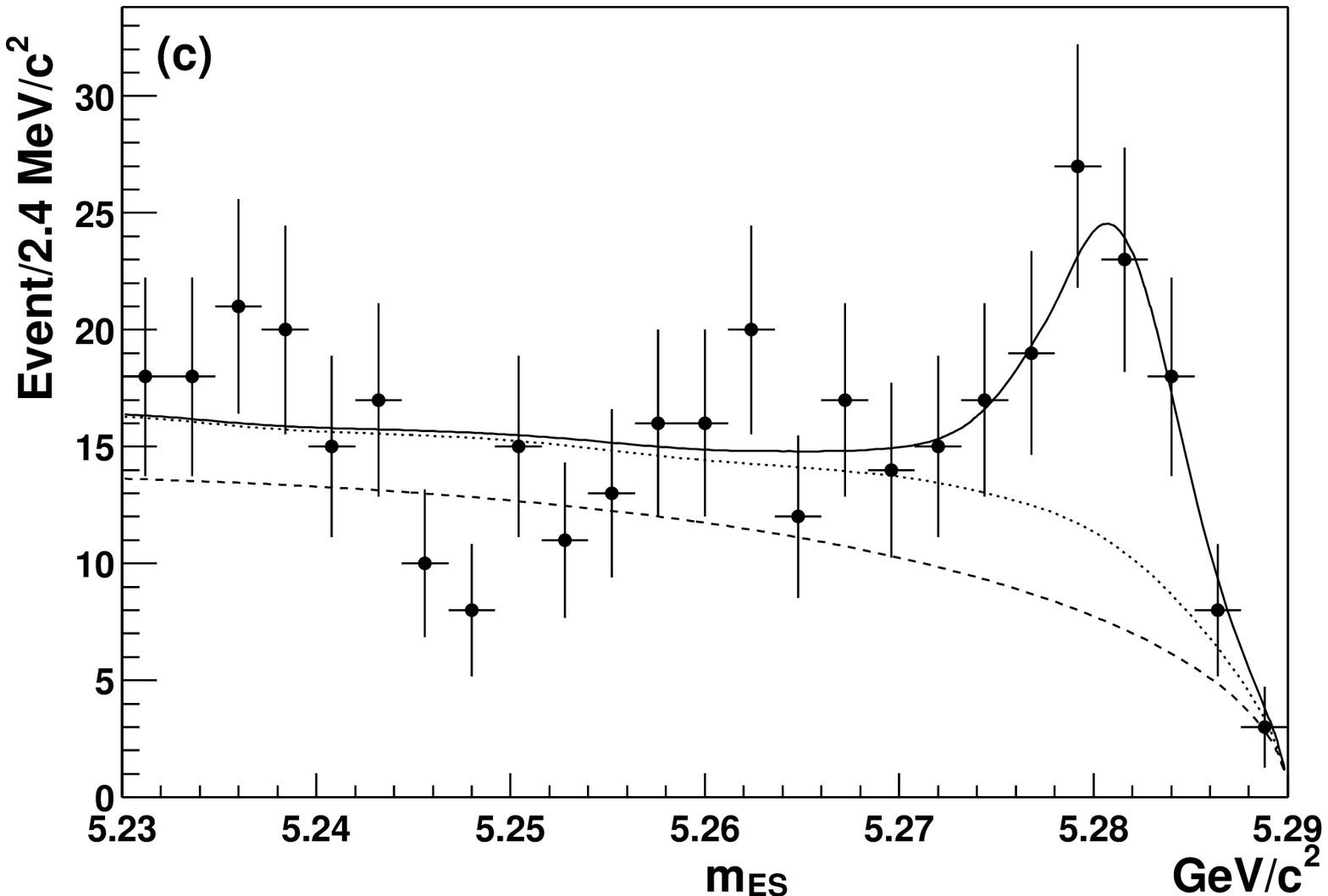}
                \epsfxsize4.5cm\epsffile{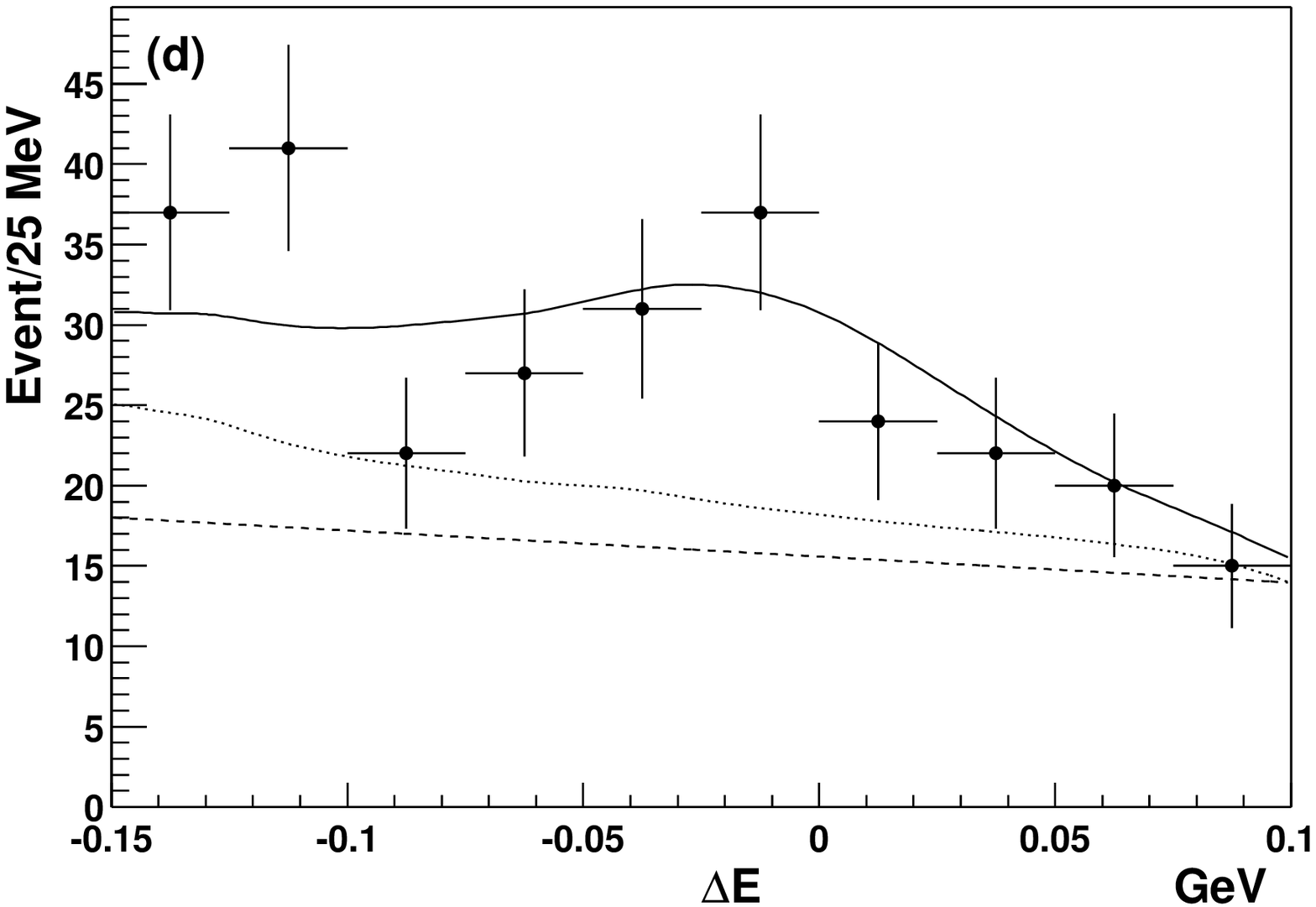}} 
  \centerline{ 	\epsfxsize4.5cm\epsffile{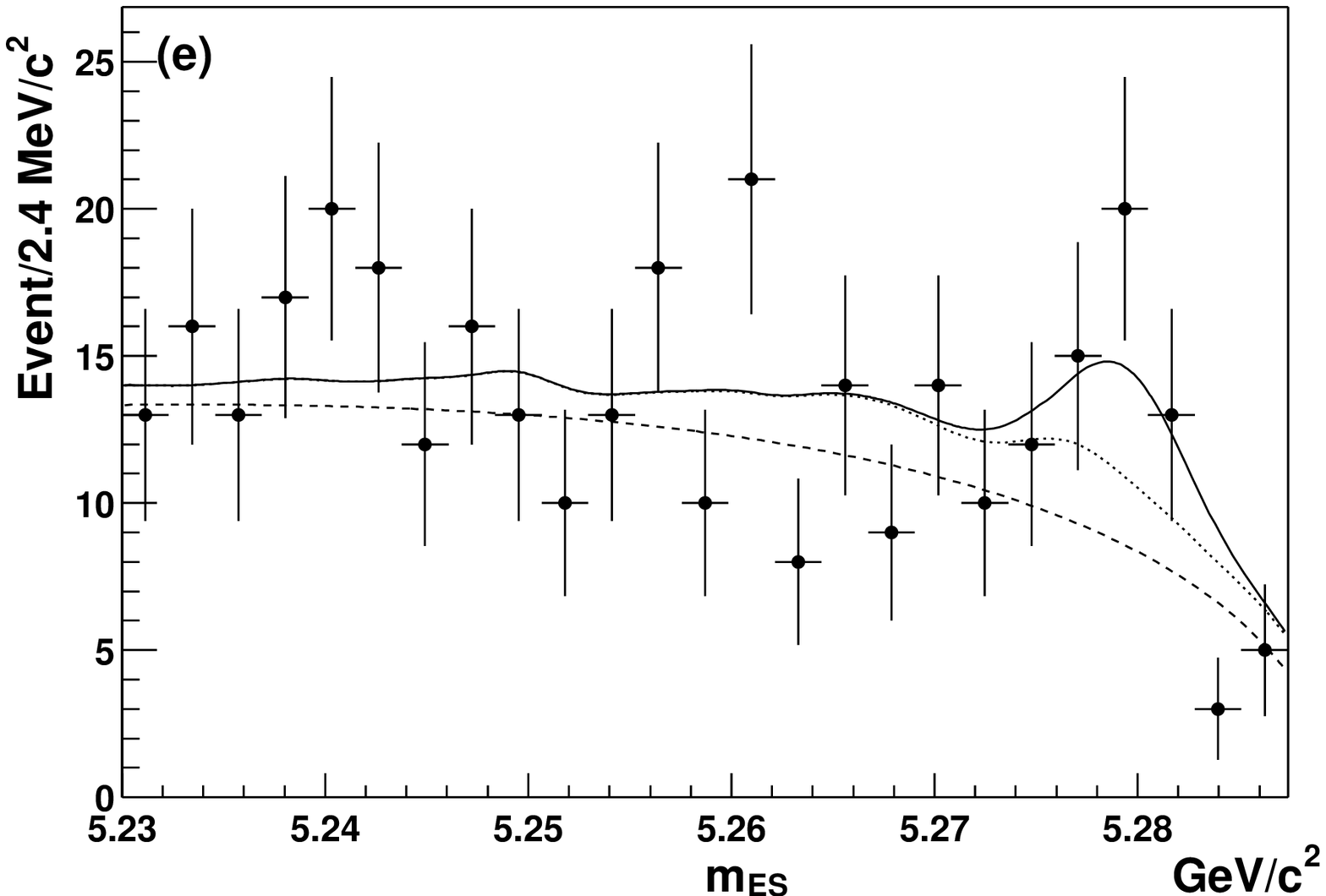}
                \epsfxsize4.5cm\epsffile{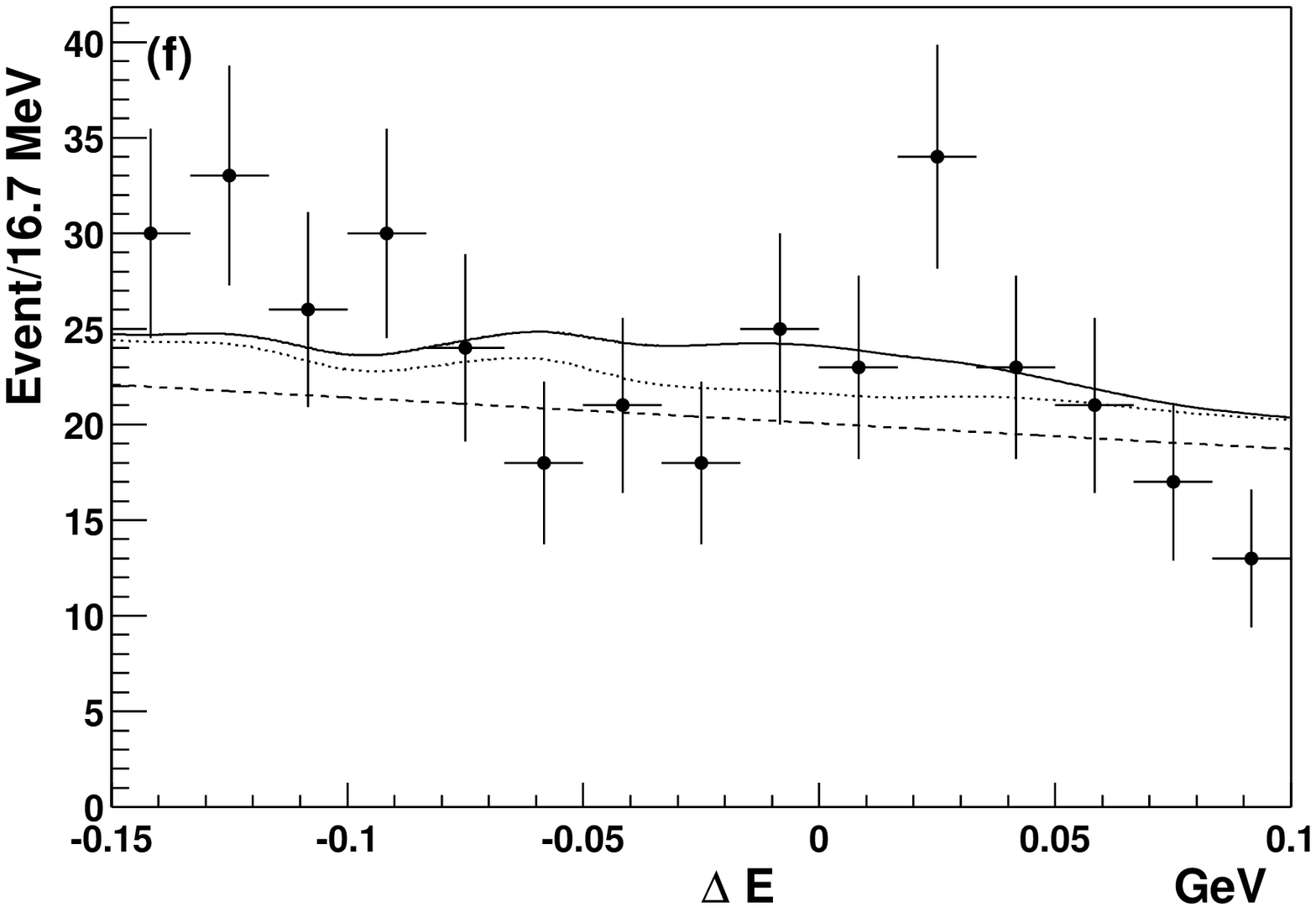}} 
\vspace{-0.3cm}
\caption{Distributions of $\mes$ and $\dE$ for samples enhanced 
	in $\rpc$ signal (a,b), $\rpn$ signal (c,d) and $\rzp$ signal (e,f).
	The solid curve represents a projection of the maximum 
	likelihood fit result. The dashed curve represents the contribution 
	from continuum events, and the dotted line indicates the combined 
	contributions from continuum events and \B-related backgrounds.}
\label{fig:ProjMesDE}
\end{figure}
After correcting for the fit biases we find from the maximum likelihood fits 
the event yields,
$
N({\rho^+\pi^0}) = 169.0\pm28.7$, 
$N({\rho^0 \pi^+}) = 237.9\pm26.5$,
and $N({\rho^0 \pi^0}) = 24.9\pm11.5$,
where the errors are statistical only. Figure~\ref{fig:ProjMesDE} shows 
distributions of $\mes$ and $\dE$, 
enhanced in signal content by cuts on the signal-to-continuum 
likelihood ratios of the other discriminating variables. 
The statistical significance of the previously unobserved $\bchtorchp$ signal
amounts to $7.3 \sigma$,  computed as 
$\sqrt{2\Delta \mathrm{log}{\cal L}}$, where $\Delta$log${\cal L}$ is the 
log-likelihood
difference between a signal hypothesis corresponding to the bias-corrected 
yield and a signal hypothesis corresponding
to a yield that equals one standard deviation of the systematic error.
We find the branching fractions to be
\begin{eqnarray*}
  {\cal B}(\B^+\to\rho^{+}\pi^{0}) &=& 
        (10.9 \pm 1.9 \pm1.9)\times 10^{-6}\,, \\
  {\cal B}(\B^+\to\rho^{0}\pi^{+}) &=& 
        \phantom{1}(9.5 \pm 1.1 \pm 0.8)\times 10^{-6}\,, \\
  {\cal B}( \B^0\rightarrow\rho^0\pi^0)  &=&
	\phantom{1}(1.4 \pm 0.6 \pm 0.3)\times 10^{-6}\,, 
\end{eqnarray*}
where the first errors are statistical and the second systematic.
The systematic errors include the uncertainties in the efficiencies, 
which are dominated by the uncertainty in the $\pi^0$ reconstruction 
efficiency and in the case of $\rho^0\pi^+$, by the uncertainty due to 
particle identification.

Here we define the $B^0\to\rho^0\pi^0$ branching ratio by including 
those events that pass our selection and are fitted as signal but 
excluding those events that can be interpreted as $B^0\to\rho^+\pi^-$ 
with a $\rho^+$, whose
mass is closer to $0.77\gevcc$ than the mass of the reconstructed $\rho^0$.
The signal significance for $\rho^0 \pi^0$, including statistical and 
systematic errors, is $2.1 \sigma$, and we use 
a limit setting procedure similar to Ref.~\cite{Frequentist} to obtain a 
$90\%$ Confidence-Level upper limit on its branching fraction. 
Fits on MC samples are used to find the signal hypothesis for which
the ratio of the probablity that the fitted signal yield is less than
that observed in data, and the probablity that the fitted yield is
less than that in data under the null signal hypothesis, is 0.1.
This signal hypothesis is shifted up by one sigma of the systematic error
and the efficiency is shifted down also by one sigma.
This method gives an upper limit of 
${\cal B}( \B^0\rightarrow\rho^0\pi^0) < 2.9\times 10^{-6}$.

Theoretical predictions of the ratio of branching fractions
$R\equiv {\cal B}(\Bz\to\rho^{\pm}\pi^{\mp})/{\cal B}(\B^{+}\to\rho^{0}\pi^{+})$,
vary over a wide range. 
Tree level estimates suggest $R\simeq 6$~\cite{bib:Bauer},
while the inclusion of penguin contributions,
off-shell $B^*$ excited states and scalar $\pi^+\pi^-$ resonances
leads to lower values, $R\simeq2-3$~\cite{bib:Gardner}.
Using the measured $\bchtorpch$ branching fraction and the 
$B^0\to\rho^{\pm}\pi^{\mp}$ branching fraction from Ref.~\cite{bib:350PRL}
we find $R=2.38^{+0.37}_{-0.31}({\rm stat})^{+0.24}_{-0.20}({\rm syst})$, 
which is in agreement with previous experimental
results~\cite{bib:cleo_rhopi}.

For the charged \B decays we find the charge asymmetries,
$ A^{\rho^{+}\pi^0}_{CP} = 0.24 \pm 0.16 \pm 0.06, 
A^{\rho^{0}\pi^{+}}_{CP} = -0.19 \pm 0.11 \pm0.02$,
with contributions to the systematic errors listed in 
Table~\ref{tab:sys_table}.

In summary, we have presented measurements of branching 
fractions and \CP-violating charge asymmetries in $\bchtorchp$ and 
$\bchtorpch$ decays, and a search for the decay $\btorp$.
We observe the decay $\B^{+}\rightarrow\rho^{+}\pi^0$ with a statistical 
significance of $ 7.3\sigma$.
We also find a branching fraction for $\B^{+}\to\rho^0\pi^{+}$ that is 
consistent with previous measurements~\cite{bib:cleo_rhopi}, 
and set an upper limit for $\btorp$. We 
do not observe evidence for direct \CP violation.

\par

We are grateful for the excellent luminosity and machine conditions
provided by our \pep2\ colleagues, 
and for the substantial dedicated effort from
the computing organizations that support \babar.
The collaborating institutions wish to thank 
SLAC for its support and kind hospitality. 
This work is supported by
DOE
and NSF (USA),
NSERC (Canada),
IHEP (China),
CEA and
CNRS-IN2P3
(France),
BMBF and DFG
(Germany),
INFN (Italy),
FOM (The Netherlands),
NFR (Norway),
MIST (Russia), and
PPARC (United Kingdom). 
Individuals have received support from the 
A.~P.~Sloan Foundation, 
Research Corporation,
and Alexander von Humboldt Foundation.


\begin{thebibliography}{99}


\bibitem{bib:350PRL}    \babar\ Collaboration, B. Aubert \ea, 
                        hep-ex/0306030,
                        submitted to Phys. Rev. Lett. (2003).

\bibitem{unitarity}	N.~Cabibbo, \jprl{10}, 531 (1963); 
                        M.~Kobayashi, T. Maskawa, 
                        \progtp{49}, 652 (1973).

\bibitem{bib:Nir}       H.J.~Lipkin, Y.~Nir, H.R.~Quinn and A.~Snyder,
                        Phys. Rev. {\bf D44}, 1454 (1991).

\bibitem{footn}		If not otherwise stated, charge conjugate modes are 
			implied throughout this document.

\bibitem{bib:cleo_rhopi}        
                        CLEO Collaboration, C.P.~Jessop {\em et al.}, 
                        \jprl{85}, 2881 (2000);
			Belle Collaboration, A. Gordon et al, 
			Phys. Lett. {\bf B542}, 183 (2002).

\bibitem{roy}		R.~Aleksan, I.~Dunietz, B.~Kayser and F. Le ~Diberder,
                        Nucl. Phys. {\bf B361}, 141 (1991).


\bibitem{bib:babarNim}  \babar\ Collaboration, B.~Aubert {\em et al.},
                        Nucl.\ Instrum.\ Methods {\bf A479}, 1 (2002).

\bibitem{bib:babarsin2b}\babar\ Collaboration, B.~Aubert {\em et al.}, 
                        Phys. Rev.~{\bf D66}, 032003 (2002).


\bibitem{PDG}           Particle Data Group, K.~Hagiwara \ea, 
                        Phys. Rev. {\bf D66}, 010001 (2002).


\bibitem{bib:argusshape}ARGUS Collaboration, H.~Albrecht {\em  et al.}, 
                        Z. Phys. {\bf C48}, 543 (1990).


\bibitem{Frequentist}   ALEPH, DELPHI, L3 and OPAL Collaborations, the LEP
                        working group for Higgs boson searches,
                        CERN-EP/98-046 (1998).

\bibitem{bib:Bauer}	M. Bauer, B. Stech and M. Wirbel,  
			Z. Phys. {\bf C34}, 103 (1987).
                   
\bibitem{bib:Gardner}   A. Deandrea and A. D. Polosa, 
			Phys. Rev. Lett. {\bf 86}, 216(2001);
			J. Tandean and S. Gardner  
			Phys. Rev. {\bf D66}, 034019(2002);
			S. Gardner and U.-G. Mei$\beta$ner 
			Phys. Rev. {\bf D65}, 094004(2002).



\end{thebibliography}
\end{document}